\documentclass[12pt,preprint]{aastex}

\slugcomment{Accepted for Publication in ApJ Letters}

\begin{document}

\title{
The First Measurement of Spectral Lines in a Short-Period 
Star Bound to the Galaxy's Central Black Hole:  A Paradox of Youth}

\author{A. M. Ghez\altaffilmark{1,2}, G. Duch\^ene\altaffilmark{1}, 
K. Matthews\altaffilmark{3}, S. D. Hornstein\altaffilmark{1}, 
A. Tanner\altaffilmark{1}, J. Larkin\altaffilmark{1}, M. Morris\altaffilmark{1},
E. E. Becklin\altaffilmark{1}, S. Salim\altaffilmark{1}, 
T. Kremenek\altaffilmark{1,4}, D. Thompson\altaffilmark{3}, 
B. T. Soifer\altaffilmark{3}, G. Neugebauer\altaffilmark{3,5}, 
I. McLean\altaffilmark{1}}

\altaffiltext{1}{Department of Physics and Astronomy, University of California, Los Angeles, CA 90095-1562; ghez, duchene, seth, tanner, larkin, morris, becklin, samir, kremenek, mclean@astro.ucla.edu}
\altaffiltext{2}{Institute of Geophysics and Planetary Physics, University of California, Los Angeles, CA 90095-1565}
\altaffiltext{3}{Caltech Optical Observatories, California Institute of Technology, MS 320-47, Pasadena, CA 91125; kym, djt, bts, gxn@irastro.caltech.edu}
\altaffiltext{4}{Department of Computer Science, Stanford, Palo Alto CA }
\altaffiltext{5}{Steward Observatory, University of Arizona, Tucson, AZ 85721}

\begin{abstract}
%With the introduction of an adaptive-optics-fed spectrometer on the 
%W. M. Keck 10 m telescope, 
We have obtained the first detection of spectral 
absorption lines in one of the high-velocity stars in the vicinity of 
the Galaxy's central supermassive black hole.
Both Br $\gamma$ (2.1661 $\mu$m) and He I (2.1126 $\mu$m) are seen in
absorption in S0-2 with equivalent widths (2.8 $\pm$ 0.3 \AA$~$ \&
1.7 $\pm$ 0.4 \AA) and
an inferred stellar rotational velocity (220$\pm$40~km/s) that are consistent 
with that
of an O8-B0 dwarf, which suggests that it is 
a massive ($\sim$15 $M_{\odot}$), young ($<$10 Myr) main sequence star.  
This presents a major challenge to star formation theories, given the 
strong tidal forces that prevail over all distances reached by S0-2 in its
current orbit (130 - 1900 AU) and
the difficulty in migrating this star inward during its lifetime
from further out where
tidal forces should no longer preclude star formation.
The radial velocity measurements 
($< v_z >$ = -510$\pm$40~km/s)
and 
our reported proper motions for S0-2 strongly constrain its orbit, providing
a direct measure of the black hole mass of 
$4.1 ( \pm 0.6 ) \times 10^6 ((\frac{R_o}{8kpc})^3 M_{\odot}$.
The Keplerian orbit parameters have uncertainities that are reduced
by a factor of 2-3
compared to previously reported values and include, for the first time, an 
independent solution for the dynamical center; this location, while consistent
with the nominal infrared position of Sgr A*, is localized to a factor 
of 5 more precisely ($\pm$2 milli-arcsec).
Furthermore, the ambiguity in the 
inclination of the orbit is resolved with the addition of the radial velocity
measurement, indicating that the star 
is behind the black hole at the time of closest approach and 
counter-revolving against the Galaxy.
With further radial velocity measurements in the
next few years, the orbit of S0-2 will provide the 
most robust estimate of the distance to the Galactic Center.

\end{abstract}

\keywords{black hole physics -- Galaxy:center --- Galaxy:kinematics and dynamics --- 
infrared:stars -- techniques:high angular resolution --- 
techniques:spectroscopic}  

\section{Introduction}

Ten-meter class ground-based telescopes
present the opportunity to obtain an unprecedented
view of the Galactic Center in terms of both sensitivity
and angular resolution.  Initial studies of the Galaxy's central
cluster at high angular resolution relied upon speckle imaging
techniques.  This first round of experiments was able to measure
stellar motions on the plane of the sky, yielding estimates of the
projected velocities (Eckart \& Genzel 1996; Ghez et al. 1998), 
projected accelerations (Ghez et al.  2000; Eckart et al. 2002), and  
three-dimensional orbital motions 
(Sch\"odel et al. 2002; Ghez et al. 2003), which each provided a successively 
stronger case for a supermassive
black hole at the center of the Milky Way and its association with the
unusual radio source Sgr A* (Lo et al. 1985).

The advent of Adaptive Optics (AO) makes it practical to obtain
spectroscopic observations of individual stars in the central 
$1 {\tt''} \times  1 {\tt''}$ of the Galaxy.  Such measurements
in principle provide the critical third dimension of motion, as well as 
astrophysical information about these high velocity stars, such as 
spectral type and rotational velocity.
While speckle spectroscopy was attempted by Genzel et al. (1997), it was 
limited in spectral resolution to R$\sim$35 and by source confusion.  The first
AO-assisted spectroscopy of the Galactic Center was attempted with 
significantly higher spectral resolution, 
R$\sim$2,000 (Gezari et al 2002).  
Nonetheless, neither of these
experiments yielded any line detections in the high velocity ($>$1000 km/s)
stars, 
indicating that the spectral lines
are weak and would require either even higher spectral resolution or higher
SNR (or possibly both) to be detected.

In this paper we report the first measurement of spectral absorption 
lines in one of the high velocity stars, S0-2.
Section 2 describes the observations, 
Section 3 details the results, 
and Section 4 discusses
how these observations improve the orbital solutions
and raises questions regarding 
the origin of this apparently massive, young star.

\section{Observations and Data Analysis}

On the nights of 2002 June 2-3 (UT), near-infrared spectra of S0-2
were obtained with the W. M. Keck 10 m telescope's Adaptive
Optics system (Wizinowich et al. 2000) and NIRC2, the facility 
near-infrared instrument (Matthews et al., in prep).   AO corrections were 
made with a R=13.2 mag natural guide star (USNO 0600-28579500) located 30" 
away from S0-2.  Using NIRC2 with its medium resolution grism, 2 pixel slit, 
and medium resolution pixel scale of $\sim$0\farcs 02/pixel, we achieved 
a spectral resolution of R$\sim$4000 ($\sim$75 km/s) and a spectral range
of 0.3 $\mu m$ across the 1024x1024 array.  On the first night, the 
K-bandpass
filter was used with the slit positioned such that the 
spectral coverage was 2.04--2.30\,$\mu$m.  Seven 20 min exposures were
obtained with S0-2 at various locations on the slit.   HD~195500
(spectral type A1V) and HD~193193 (spectral type G2V) were
observed as calibration sources at the
same locations on the slit as S0-2.  On the second night,
the setup was altered somewhat to help identify and minimize
systematic effects, such as fringing, which might complicate any
line detections.  The filter was changed to the K'-bandpass,
the slit was moved, giving a spectral coverage of
2.03-2.29\,$\mu$m, and the calibration sources observed were
HD~190285 (spectral type A0V) and HD~193193. Four spectra of
S0-2 were obtained with this set-up.  Because of the high stellar densities
at S0-2's location, a
dark patch of sky 1$^{hr}$ East of Sgr A* was observed
on both nights to measure a purely sky background.
The wavelength solution was
obtained by identifying a set of 16 OH emission lines in the
spectra of the sky and by fitting a low-order polynomial
function to the location of those lines throughout the
detector along the spatial direction. The accuracy of this
process is $\sim9$\,km/s, as measured by the dispersion of the residual
of the fit. 

The sky-subtracted spectrum of S0-2 in each dataset was
extracted over a 0\farcs 06 spatial interval along the slit and divided by the spectrum
of the A-type star, extracted over the same window, to
properly correct for the instrumental spectral response.  Prior
to this step, the spectrum of the G-type star was used to
remove the strong Br$\gamma$ absorption feature in the
spectrum of the A-type star (cf. Hanson et al. 1996). 
The spectrum was then divided by a
blackbody of the temperature matching the spectral type
of the calibrator to provide a spectrum corrected for all
telluric absorption features.

The extracted spectra of S0-2 were still partially
contaminated by background emission due to the gas present
around the Galactic Center as well as to the presence of the
nearby bright source IRS~16C, whose wings 
extend along the spatial direction 
up to the location of S0-2. To correct for
this emission, we estimated a local background by averaging
the flux in two 0\farcs 06 spatial intervals along the slit
located 0\farcs75 to the SW
and 0\farcs4 to the NE of S0-2. This background 
estimate, which
amounts to 40--50\,\% of the level of S0-2's continuum, was subtracted from the
spectrum of the source to provide the final spectrum, which  
has SNR per pixel of $\sim$30-80 in the continuum 
%\footnote{The exact SNR depends on the wavelength, with the locations of
%telluric absorption features having lower SNR} 
(see Figure 1).  
There is significant Br$\gamma$ emission from the
gas surrounding the Galactic Center and the intensity of this
line varies spatially. Close to S0-2, however, the wavelength
of this feature does not vary, which helps to distinguish it
from any potential photospheric feature. 
In general, as long as the magnitude of S0-2's Doppler shift is greater than
$\sim$300 km/s, the line emission from the gas 
is unlikely to cause problems, either non-detections or biases, for 
R$\sim$ 4000 measurements of stellar Br $\gamma$ absorption.  The zone of gas
contamination is estimated based on the structure of the Br$\gamma$ emission
line detected in the gas close to S0-2.  The residual emission
feature seen in the final spectrum of S0-2 corresponds to this local gas
emission, and we believe that it is not physically associated with the star
(see \S3).

\section{Results}

The spectrum of S0-2, shown in Figure 1, 
has two identifiable spectral lines.
These are both seen in absorption and are identified as
the H I (4-7) or Br$\gamma$ line at 2.1661 $\mu$m and the He I
triplet at 2.1126 $\mu$m ($3p ~ ^3P^o-4s ~ ^3S$), which is a blend of 
three transitions at 2.11274, 2.11267, and 2.11258 $\mu$m.
Table~1 summarizes the properties of these two lines, which are
obtained by fitting the background continuum over the whole
spectrum with a low-order polynomial and fitting the lines with a
Gaussian profile.  Also given in Table 1 are limits on three lines that 
are not detected.  
The reported values and uncertainties are
based on the fits to the average of all 11 spectra and the standard
deviation of the mean of the fits to 5 independent pairs of spectra.
The locations of the Gaussian peaks provide estimates of S0-2's Doppler
shift, after a correction of 18 km/s is applied to account for the Earth's 
motion around the Sun and the Sun's motion towards the center of the Galaxy (Binney \& Merrifeild 1998).
The agreement between the Doppler shifts inferred
from the Br$\gamma$ and He I absportion line profiles supports the 
interpretation of the Br$\gamma$ emission in S0-2's spectrum as residual gas emission as opposed to part of a P Cygni line profile.
The weighted average of these two Doppler shift measurements 
yields a radial velocity of
-510$\pm$40~km/s for S0-2 during 2002.4187.  
Combining this radial velocity with the tangential velocity measured
by Ghez et al. (2003) at the same epoch, results in a total space 
motion of 6660 $\pm$ 730 km/s.

The Gaussian half width at half maximum (HWHM) values provide estimates of 
the rotational
velocities. The intrinsic rotational profile of a star has a
HWHM that corresponds to 90\% of $v\sin i$ (Gray 1976). Since
the HWHM of both lines is significantly larger than the
instrumental HWHM ($\sim$38 km/s), we simply scale the Gaussian HWHM 
values by
a factor of 1.1 to derive an estimated value of $v\sin i$.
The values from the two lines are comparable and averaged
together suggest a $v\sin i$ of 220$\pm$40~km/s.

Equivalent widths (EW) for the lines are also derived from the
Gaussian fits.  The EW(Br$\gamma$) is 2.8 $\pm$ 0.3 \AA$~$and
the EW(HeI[2.1126 $\mu$m]) is 1.7 $\pm$ 0.4 \AA. 
These values are checked against those obtained from
integrating over 40\,\AA-windows centered on the lines, which
provide similar results to within 3--7\%. 
Using the same window and the Doppler shift measured for two detected
lines, we derive the limits for He I (2.0581 $\mu$m),
N III (2.1155 $\mu$m), and He II (2.1885 $\mu$m), which are listed
in Table 1.

Between our two consecutive nights of observations
there is an expected change of 60 km/s in the radial velocity (\S4.1). 
We therefore have also analyzed the spectra averaged over each
night separately.  The results for the Br$\gamma$ line
are reported in Table 1.
While the change in radial velocities between the two nights
is not significant ($V_z$(2002.4205) - $V_z$(2002.4177) = -37 $\pm$ 47 km/s), 
it is consistent with the predicted change.  
Furthermore, the EW and $v\sin i$ measurements from the spectrum averaged 
over the two nights appears to be unaffected by this shift.
The individual nights' $V_r$ are used in the
dynamical analysis in \S4.1 and the averaged spectrum values of
EW and $v\sin i$ are applied in the stellar astrophysics discussion
in \S4.2.

\section{Discussion \& Conclusions}

\subsection{Dynamics}

The measurements of a radial velocity for S0-2 provide a new
and powerful constraint on its orbit.   S0-2's
motion on the plane of the sky provided the first estimates
of the orbital parameters (Sch\"odel et al. 2002; Ghez et al. 2003).
By combining the radial velocity
with the proper motions reported by Ghez et al. and 
assuming a distance of 8.0 kpc (Reid 1993), as was done in both 
proper motion analyses,  we obtain the first orbital solution for S0-2
based on measured three dimensional motion  
(Table 2).  
We assume here that Sgr A* has no significant velocity with 
respect to the Galaxy along the line of sight, as is observed for its
proper motion (Reid et al. 1999).
Our fit produces a total mass estimate of 
$4.1 ( \pm 0.6) \times 10^6 (D/8kpc)^3 
M_{\odot}$, consistent at the $\sim$2$\sigma$ level with
earlier estimates based on velocity dispersion measurements.
Our results on S0-2's orbit do not assume the location of the dynamical
center, as opposed to the analysis of Sch\"odel et al..  Despite these
two additional free parameters, the uncertainties on the orbital
parameters are reduced by a factor of 2-3.  This first orbital 
estimate of the Galaxy's
dynamical center is not only consistent with the nominal 
infrared position of Sgr A* to within the uncertainties on the latter 
(Reid et al. 2003), but also a factor
of 5 more precise ($\pm$ 2 milli-arcsec).

The addition of radial velocity measurements also breaks the 
ambiguity in the inclination angle, {\it i}.  
With the proper motion data alone, only the 
absolute value of the inclination angle can be determined,
leaving the questions of the direction of revolution and
where along the line of sight the star is located behind the black hole 
unresolved.
Our radial velocity measurements indicate a negative inclination angle
and consequently that S0-2 is both counter-revolving against the Galaxy
and behind the black hole at the time of periapse.
The improved location of the center of attraction 
%($\pm$ 2 milli-arcsec) 
from the orbital analysis 
results in a minimum offset of S0-2 from the blackhole 
in the plane of the sky of 11 $\pm$ 2 milli-arsec, which is 
significantly larger than the expected Einstein radius 
($\theta_E$ = 0.42 milli-arcsec for the S0-2 distance behind the black hole
of $\sim$ 100 AU) and therefore makes
gravitational lensing an unlikely event (Wardle \& Yusef-Zadeh 1992; 
Alexander \& Loeb 2001).  

In principle, the addition of radial velocities to the study of S0-2's dynamics
allows the distance to the Galactic Center, $R_o$, to be a free parameter in 
the orbital fits (Salim \& Gould 1999).  The measurements, however, were 
obtained just 30 days after the star's 
closest approach to the black hole when the radial velocity was changing very
rapidly (see Figure 2).  
While the current radial velocity and proper motion data set constrains 
$M/{R_o}^3$ very effectively ($\sim$15\% uncertainty), it does not yet 
produce a meaningful measurement
of $R_o$.  Nonetheless, as Figure 2 shows, the radial velocities 
from the currently allowed orbits quickly diverge, producing a spread
of a few hundred km/s in one year.
%\footnote{
%In 2000.47, the epoch of the earlier AO spectroscopic observations reported by 
%Gezari et al. (2002), the radial velocity is predicted 
%to be $\sim$1240 km/s, well outside the region
%affected by the gas line emission.  Therefore the non-detection 
% is entirely attributable to a factor of 3 lower SNR
%per equivalent spectral resolution as opposed to the lower spectral 
%resolution of R$\sim$2000.  This drop in sensitivity was caused by
%the external optics used to convert a 
%non-AO instrument into one that could be used as an AO instrument in the
%interim before the AO designed instruments were ready. }
Within the next few years,  the orbital fits based on both proper motions
and additional radial velocity measurements should provide the most 
direct and precise estimate of the distance to the Galactic Center, making
it a fundamental rung in the cosmic distance ladder.

%With a period of a mere 15 yrs, S0-2 opens up an aditional new realm for 
%dynamical studies in the Galactic Center;  it offers a  
%unique opportunity to look for deviations from a Keplerian orbit.
%These might arise from precession of the periapse distance due
%to general relativistic effects (Fragile \& Mathews 2000) or an extended 
%mass distribution (Rubilar \& Eckart 2001), in the form of either an entourage 
%of stellar remnants surrounding the central supermassive black hole,
%a spike of dark matter particles (Gondolo \& Silk 1999; Ullio et al. 2001)
% or a binary black hole.

\subsection{Stellar Astrophysics}

The detection of absorption lines allows us to sort out the spectral
classification ambiguities present when only photometric information
is available and to determine if this star's photosphere has been
altered as a result of its close proximity to the central black hole.  
The average brightness at 2.2 $\mu$m for S0-2 is K $\sim$ 13.9  
mag and there is no evidence of brightening after periapse passage
(Ghez et al. 2003).  With a distance of 8.0 kpc and 
K-band extinction of 3.3 mag (Rieke, Rieke, \& Paul 1989), 
the 2.2 $\mu$m brightness of S0-2 implies that, if it is an ordinary star
unaltered by its environment, it could either be an O9 main-sequence star 
or a K5 giant star;
all supergiants are ruled out as they are too bright by at least 2 magnitudes
in the K bandpass.
Kleinmann and Hall (1986) provide a 2.0 - 2.5 $\mu$m spectral atlas of 
late-type stars that demonstrates that if S0-2 is a K5 giant star, then it
should have deep CO absorption lines, which definitively were not detected
in either this experiment or our earlier
experiment reported by Gezari et al. (2002).  In contrast, 
the spectral atlas of 180 O and B stars constructed by Hanson, Conti and
Rieke (1996) shows that an O9 main sequence star both lacks the 
CO absorption and has Br $\gamma$ and He I (2.1126 $\mu$m) consistent 
with the observed values.  
% need to address 2.058 HeI and 2.18 HeII
Futhermore, stars in this comparison sample 
earlier than O8 show NIII (2.115 $\mu$m) in emission and 
He II (2.1885 $\mu$m) in absorption above our 3 $\sigma$ thresholds, which 
are listed in Table 1;  the lack of photospheric He I (2.058 $\mu$m) absorption 
does not provide any additional constraints. 
Similarly, a dwarf B-type stars later than B0 
have absorption equivalent widths that are too large.
Together, the photometry
and absorption line-equivalent widths permit dwarf spectral types ranging
from O8 to B0.  Likewise, the rotational velocity of 224 km/s is reasonable for
this range (Gatheier, Lamers, \& Snow 1981).
% and is equivalent to XXX\%
%of break-up velocity for an O9 star.  
S0-2, therefore,
appears to have a spectral type, and hence temperature ($\sim$30,000 K),
as well as luminosity ($\sim$10$^3$ $L_{\odot}$) that are consistent
 with a main sequence star having a mass of 
$\sim$15 $M_{\odot}$ and an age $<$10 Myr.
% XXX need a reference here

It is challenging to explain the presence of such a young star
in close proximity to a supermassive black hole.  
Assuming that the black hole has not significantly affected
S0-2's appearance or evolution, S0-2 must be younger than 
10 Myr and thus formed relatively recently.   If it has not 
experienced significant orbital evolution, its apoapse distance
of 1900 AU implies that star formation is possible in spite of
the tremendous tidal forces presented by the black hole, which
is highly unlikely.  If the star formed at larger distances from the 
black hole and migrated inward, then the migration would have to be through a 
very efficient process.  Current understanding of the distribution of stars,
however, does not permit such efficient migration.
This problem is similar to that raised by the He I emission-line stars
(e.g., Sanders 1992,1998;
Morris 1993, Morris et al. 1999; Gerhard 2001; Kim \& Morris 2003),
which are also counter-revolving against the Galaxy (Genzel et al. 1997),
but amplifies it with a distance from the black hole that is an order of
magnitude smaller.
An alternative explanation for S0-2's hot photosphere is that it
 may be significantly altered by its environment.
% XXXX should raise effects of passages through the black hole's accretion
% disk here
While its periapse passage is too large for it to be tidally heated
by the black hole as explored by Alexander \& Morris (2003), it may be
affected by the high stellar densities found in this region.  
On the one hand, the high stellar densities might allow S0-2 to be 
an older giant star that has had
its outer atmosphere stripped through collisions; however, to generate
the necessary luminosity,  significant external heating is required 
(Alexander 1999).  On the other hand, high stellar densities might 
lead a cascade of merger events (Lee 1996), which 
would allow S0-2's formation process to have begun more than
10 Myr ago.   
However a large number of collisions would have had to occur to 
provide the necessary lifetime to bring it in from sufficiently
large radii.
More exotically, it could be a "reborn" 
star, which occurs as the product of a merger of a stellar remnant with 
a normal star.   None of these possibilities is altogether satisfactory, 
leaving S0-2 as a paradox of apparent youth in the vicinity of 
a supermassive black hole.
% should we reference a more complete discussion of this in the
% other paper?

%\subsection{Conclusions}

%Using the newly-commissioned near-infrared AO instrument, NIRC2,
%on the W. M. Keck II 10 m telescope, we detected, for the first time,
%photospheric absorption lines in one of the high-velocity stars in the Galaxy's
%central cluster.   Its spectral and photometric properties are consistent with
%that of a late-O early-B dwarf.
%If this star is unaffected by its surroundings,
%these characteristics suggest that it is both
%massive, $\sim$15 $M_{\odot}$, and young, $<$10 $Myr$.  The apparent
%youth of this star is problematic given, an apoapse distance of
%only 2000 AU in its current orbit.  This dramatically intensifies
%the problem of extreme youth in the proximity of a black hole
%originally raised by the presence of He I emission line stars, 
%which reside at an order of magnitude larger radii (e.g., Sanders 1992). 

%The addition of a radial velocity measurement to the proper motions 
%significantly improves the orbital analysis of S0-2.  
%Assuming a distance of 8.0 kpc to the Galactic Center,  we 
%derive orbital parameters that have a factor of 2-5 smaller uncertainties than
%previous estimates and determine that the star was behind the black hole
%from March to September of 2002 and counter-revolves against the Galaxy.  

\acknowledgments
We thank the staff of the Keck obsevatory, especially 
Randy Campbell, Grant Hill, Chuck Sorensen, David LeMignant, and 
director Fred Chaffee.
This work has been supported by the National Science Foundation through
the individual grant AST99-88397 and the Science and Technology Center for 
Adaptive Optics, managed by the University of California at Santa Cruz under 
Cooperative Agreement No. AST - 9876783.
The research of E.E.B., B.T.S. and  D.J.T. is supported by NASA.
The W.M. Keck Observatory is operated as a scientific partnership among the 
California Institute of Technology, the University of California and the 
National Aeronautics and Space Administration. The Observatory was made 
possible by the generous financial support of the W.M. Keck Foundation.

\pagebreak

\pagebreak

\begin{deluxetable}{llll}
\tablenum{1}
\footnotesize
\tablecaption{Spectral Properties of S0-2\tablenotemark{a}}
\tablehead{\colhead{}&
	   \colhead{EW (\AA)}  &
           \colhead{$V_z$ (km/s)}  &
           \colhead{$V_{rot}$ (km/s)} 
}
\startdata
Br$\gamma$ (2.1661 $\mu$m) &	  	&    	&   \nl
$~ ~ ~$ Avg. Spectrum &	2.8$\pm$0.3  &   -510$\pm$39  &  238$\pm$62 \nl
$~ ~ ~$ June 2 Spectrum &	\nodata  &   -495$\pm$36  &  \nodata \nl
$~ ~ ~$ June 3 Spectrum &	\nodata  &   -532$\pm$44  &  \nodata \nl
He I (2.1125 $\mu$m)	   &	1.7$\pm$.4   &   -532$\pm$96  &  216$\pm$53 \nl
He II (2.1891 $\mu$m)	   &    $<$0.5         &   \nodata      &  \nodata    \nl
N III (2.1155 $\mu$m)	   &    $<$1.0         &   \nodata      &  \nodata    \nl
He I (2.0587 $\mu$m)	   &    $<$1.3	      &	  \nodata      &  \nodata    \nl
Average 		   &	\nodata 	& -513$\pm$36  &  224$\pm$40 \nl
\enddata
%\tablecomments{The uncertainties and limits listed are 1$\sigma$ and 3$\simga$ values, respectively}
\tablenotetext{a}{For all lines except Br$\gamma$, only the results from 
the averaged spectrum are listed.}
\end{deluxetable}

\begin{deluxetable}{lll}
\tablenum{2}
\footnotesize
\tablecaption{Orbital Solutions for S0-2}
\tablehead{\colhead{Orbital Parameter\tablenotemark{a}}&
	   \colhead{This Work}  &
           \colhead{Sch\"odel et al.\tablenotemark{b}} 
}
\startdata
$\Delta x_o$ (milli-arcsec) 	& -2.7 $\pm$ 1.9 		& \nodata \nl
$\Delta y_o$ (milli-arcsec) 	& -5.4 $\pm$ 1.4 		& \nodata \nl
M ($(\frac{R_o}{8kpc})^3 10^6 M_{\odot}$)	& 4.07 $\pm$ 0.62	 	&  3.7 $\pm$ 1.5 \nl
A (milli-arcsec)		& 125.6 $\pm$ 5.5   	&  119 $\pm$ 15 \nl 
P (yrs)				& 15.78 $\pm$ 0.82 	&  15 $\pm$ 1  \nl
e 				& 0.8736 $\pm$ 0.0083 	&  0.87 $\pm$ 0.03\nl
T$_o$ (yrs) 			& 2002.334 $\pm$ 0.017 	&  2002.30 $\pm$ 0.05\nl
i ($^o$)			& -47.3 $\pm$ 2.5	& $\pm$46 $\pm$ 4 \nl
$\omega$ ($^o$) 		& 248.5 $\pm$ 1.8	&  250 $\pm$ 5 \nl
$\Omega$ ($^o$) 		& 49.9 $\pm$ 3.0	&  36 $\pm$ 9 \nl
\enddata
\tablenotetext{a}{ There are nine orbital parameters included in
the fits reported here and seven in those calculated by Sch\"odel et al.
(2002), who fix the center of attraction (true focus) to the location of 
Sgr A*  reported by Reid et al. (2003).  $\Delta x_o$ and $\Delta y_o$ 
are the center of attraction's East-West and North-South offsets from
Sgr A*; the reported uncertainy in the offsets includes only the uncertainty 
in the dynamical center.  The remaining orbital parameters are semi-major axis
(A), period (P), eccentricity (e), time of periapse passage ($T_o$), inclination
(i), angle of node to periapse ($\omega$), and angle of line of nodes 
($\Omega$).  While mass (M) is not an independent parameter, it is reported here
for convenience.}
\tablenotetext{b}{ S0-2 is referred to as S2 by Sch\"odel et al. }  
\end{deluxetable}

%this uses on S0-2 to determine focus.... can do better with multiple
%orbits.... what should we be reporting? 
% need to add comment in text/table that there is one difference
% between two groups handing of the true focus

\pagebreak

\begin{figure*}
\epsscale{1.0}
\plotone{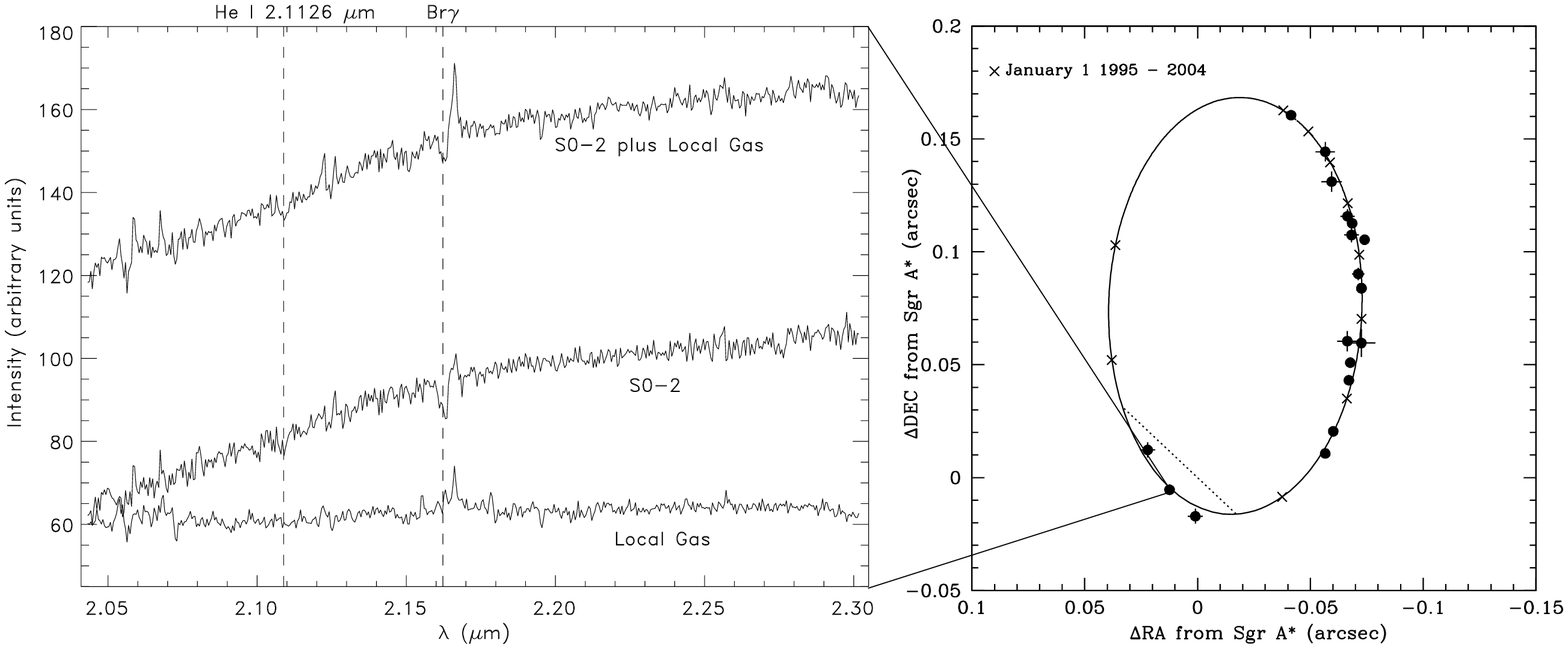}
\figcaption{  
\small In the left panel is the first spectrum of S0-2 to show detectable
photospheric absoprtion lines (Br$\gamma$ and He I (2.1126~$\mu$m)).
The final spectrum (middle) is the raw
spectrum (top; with only an instrumental background removed)
minus a local sky (bottom).  
The horizontal dimension has been re-binned by a factor of two for display 
purposes only.  
The vertical lines are drawn at 2.10899 and 2.16240 $\mu$m, 
which correspond to the locations of Br$\gamma$ and He I for
a $V_{LSR}$ of -513 km/s.
This spectrum was obtained in 2000 June at the same time as one
of the proper motion measurements reported by Ghez et al. (2003)
and shown in the right panel (filled circles).  The crosses mark January 1 of
each year between 1995 and 2004 for the best fit orbit solution 
(solid line), which
is based on both the radial velocity and proper motions.  The dotted
line is the line of nodes, which reveals S0-2 to be behind the black hole
for a mere $\sim$0.5 years out of its 15 year orbit.
}
\end{figure*}

\begin{figure}
\epsscale{1.0}
\plotone{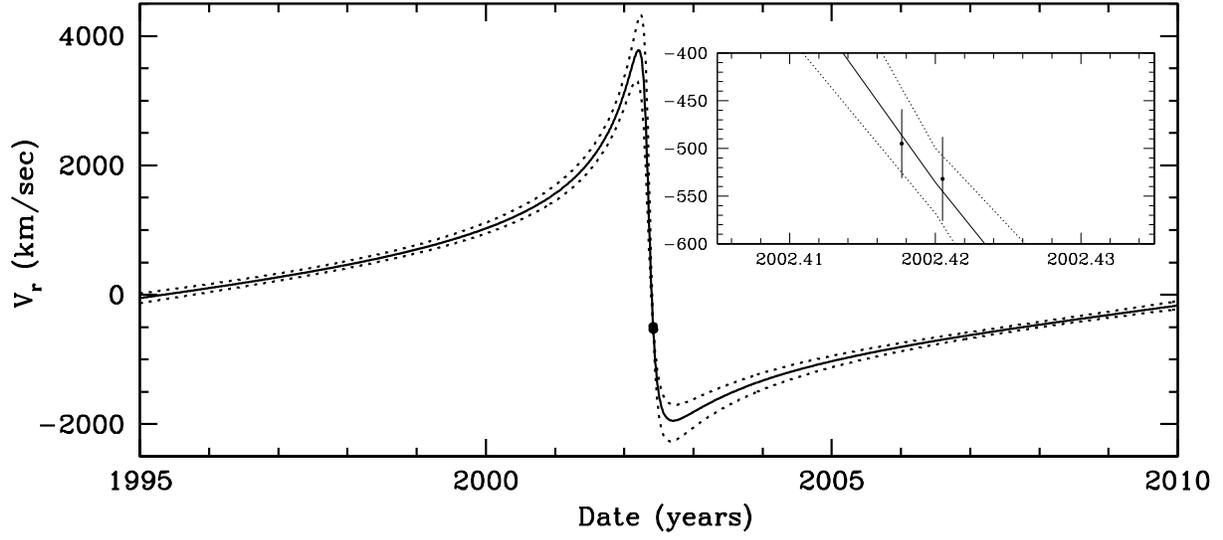}
\figcaption{ The measured radial velocity along with the predicted 
radial velocities.   The solid curve comes from the best fit orbit and the
dotted curves display the range for the orbital solutions 
allowed with the present data sets. 
}
\end{figure}


\begin{thebibliography}{} 

\bibitem[Alexander (1999)]{a99} Alexander, T. 1999, ApJ, 527, 835

\bibitem[Alexander \& Loeb (2001)]{al01} Alexander, T. \& Loeb, A. 2001, ApJ, 
551, 223                

\bibitem[Alexander \& Morris (2002)]{am02} Alexander, T., \& Morris, M. 2003,
ApJLett, submitted

\bibitem[Binney \& Merrifeild ()]{bm} Binney, J., \& Merrifield, 1998
``Galactic Astronomy," Princeton Univesity Press

\bibitem[Ecart \& Genzel (1996)]{eg96} Eckart, A., \& Genzel, R. 1996, Nature,
383, 415

\bibitem[Eckart et al. (2002)]{egos02} Eckart, A., Genzel, R., Ott, T., \&
Sch\"odel, R. 2002, MNRAS, 331, 917

%\bibitem[Fragile \& Mathews (2000)]{fm00} Fragile, P. C., \& Mathews, J. 2000, ApJ,
%542, 328

\bibitem[Gathier et al. (1981)]{gls81} Gathier, R., Lamers, H.J.G.L.M., \& Snow,
T. 1981, ApJ, 247, 173

\bibitem[Genzel et al. (1997)]{geoe97} Genzel, R., Eckart, A., Ott, T., \& Eisenhauer, F. 1997, MNRAS, 291, 219

\bibitem[Gerhard (2001)]{g01} Gerhard, O. 2001, ApJ, 546, L39

\bibitem[Gezari et al. (2002)]{g02} Gezari, S., Ghez, A. M., Becklin, E. E., 
Larkin, J., McLean, I. S., Morris, M. 2002, ApJ, 576, 790

\bibitem[Ghez et al. (2003)]{ghe03} 
Ghez, A.M., Becklin, E.E., Duchene, G., Hornstein, S., M. Morris, S. Salim,
A. Tanner (2003) Astron. Nachr., Vol. 324, No. S1, Special Supplement
"The central 300 parsecs of the Milky Way", Eds. A. Cotera, H. Falcke, T. R.
Geballe, S. Markoff

\bibitem[Ghez etal (1998)]{ghe98} Ghez, A.M., Klein, B.C., Morris, M., 
\& Becklin, E.E. 1998, ApJ, 509, 678

\bibitem[Ghez etal (2000)]{ghe00} Ghez, A.M., Morris, M., Becklin,
E.E., Tanner, A., \& Kremenek, T. 2000, Nature, 407, 349

\bibitem[Gray 1976]{g76} Gray, D. F. 1976, The Observation and Analysis of 
Stellar Photospheres (New York: Wiley-Interscience)

%\bibitem[Gondolo \& Silk (1999)]{gs99} Gondolo, P., \& Silk, J. 1999,
%Phys. Rev. Lett., 83, 1719

\bibitem[Hanson etal (1996)]{han96} Hanson, M.M., Conti, P.S.,
 \& Rieke, M.J. 1996, ApJS, 107, 281

\bibitem[Kim \& Morris (2002)]{km02} Kim, S. S., \& Morris, M. 
2003, ApJ, submitted

\bibitem[Kleinmann \& Hall (1986)]{kle86} Kleinmann, S.G., \& Hall, D.N.B
1986, ApJS, 62, 501

\bibitem[Lee 1994]{l996} Lee, H.M., 1996, IAU 169, 215 

\bibitem[Lo et al.(1985)]{1985Natur.315..124L} Lo, K.~Y., Backer, D.~C.,
Ekers, R.~D., Kellermann, K.~I., Reid, M., \& Moran, J.~M.\ 1985, \nat,
315, 124

\bibitem[Morris (1993)]{m93} Morris, M., 1993, ApJ, 408, 496

\bibitem[Morris et al. (1999)]{mgb99} Morris, M., Ghez, A. M., Becklin, E. E.
1999, Adv. Spa. Res., 23, 959

\bibitem[Reid (1993)]{rei93} Reid, M.J. 1993, ARA\&A, 31, 345   

\bibitem[Reid et al. 1999]{r99} Reid, M.J., Readhead, A.C.S., Vermeulen, 
R.C., Treuhaft, R.N. 1999, ApJ, 524, 816

\bibitem[Reid etal (2003)]{r02} Reid, M.J., Menten, K.M., Genzel, R., Ott, T., 
Sch\"odel, R., \& Eckart, A. 2003, ApJ, in press (astro-ph/0212273)

\bibitem[Rieke et al. 1989]{rrp89} Rieke, G.H., Rieke, M.J., \& Paul, A.E.
1989, ApJ, 336, 752

%\bibitem[Rubilar \& Eckart (2001)]{re01} Rubilar, G. F., \& Eckart, A. 2001,
%A\&A, 2001, 372, 95

\bibitem[Salim \& Gould (1999)]{sg99} Salim, S., \& Gould, A. 1999, ApJ, 
523, 633

\bibitem[Sanders (1992)]{s92} Sanders, R. H. 1992, Nature, 359, 131 

\bibitem[Sanders (1998)]{s98} Sanders, R. H. 1998, MNRAS, 294, 35

\bibitem[Sch\"odel et al. (2002)]{s02} Sch\"odel, R. et al. 2002, 
Nature, 419, 694

%\bibitem[Ullio et al. (2001)]{u01} Ullio, P., Zhao, H., Kamionkowski, M.
%2001, Phys, Rev. D, 64, 1302

\bibitem[Wardle etal (1992)]{wz92} Wardle, M. \& Yusef-Zadeh, F. 1992, ApJ, 
387, L65

\bibitem[Wizinowich et al. (2002)]{w00} Wizinowich, P., Acton, S. D., 
Lai, O., Gathright, J., Lupton, W., Stomski, P. 2000,
Proc. SPIE, 4007, 2

\end{thebibliography}
\end{document}